\documentclass[prd,twocolumn,nofootinbib,aps,superscriptaddress,tightenlines,preprintnumbers]{revtex4}

\usepackage{epsfig,latexsym}

\begin{document}

\preprint{\hbox{CALT-68-2626}  } 

\title{Imprints of a Primordial Preferred Direction on the Microwave Background} 

\author{Lotty Ackerman, Sean M. Carroll and Mark B. Wise}

\affiliation{California Institute of Technology, Pasadena, CA 91125 }

\begin{abstract}
Rotational invariance is a well-established feature of low-energy physics. Violations of this symmetry must be extremely small today, but could have been larger in earlier epochs. In this paper we examine the consequences of a small breaking of rotational invariance during the inflationary era when the primordial density fluctuations were generated. Assuming that a fixed-norm vector picked out a preferred direction during the inflationary era, we explore the imprint it would leave on the cosmic microwave background anisotropy, and provide explicit formulas for the expected amplitudes $\langle a_{lm}a_{l'm'}^*\rangle$ of the spherical-harmonic coefficients. We suggest that it is natural to expect that the imprint on the primordial power spectrum of a preferred spatial direction is approximately scale-invariant, and examine a simple model in which this is true.
\end{abstract}

\date\today
\maketitle

\section{Introduction}

Inflationary cosmology, originally proposed as a solution to the horizon, flatness, and monopole problems \cite{Guth:1980zm,{Linde:Albrecht}}, provides a very successful mechanism for generating primordial density perturbations.   During inflation, quantum vacuum fluctuations in a light scalar field are redshifted far outside the Hubble radius, imprinting an approximately scale-invariant spectrum of classical density perturbations \cite{quantumfluct,inflationreview}.  Models that realize this scenario have been widely discussed \cite{{Linde:1983gd},{Linde:1993cn},{Lyth:1998xn}}.  The resulting perturbations give rise to galaxy formation and temperature anisotropies in the cosmic microwave background, in excellent agreement with observation \cite{{COBE},{BOOMERANG},{ACBAR},{CBI},{VSA},{ARCHEOPS},{DASI},{MAXIMA},{WMAP3}}. 

If density perturbations do arise from inflation, they provide a unique window on physics at otherwise inaccessible energy scales.  In a typical inflationary model (although certainly not in all of them), the energy scale $E=V^{1/4}$ is related to the amplitude of density fluctuations $\delta$ and the reduced Planck mass $M_{\rm P}$ via $E \sim \sqrt{\delta}M_{\rm P}$.  Since we observe $\delta\sim 10^{-5}$, it is very plausible that inflation occurs near the scale of grand unification, and not too far from scales where quantum gravity is relevant.  Since direct experimental probes provide very few constraints on physics at such energies, it makes sense to be open-minded about what might happen during the inflationary era.  

In this paper we ask what happens when a cherished property of
low-energy physics -- rotational invariance -- is violated during
inflation.  Rotational invariance is of course a subset of Lorentz
invariance, and theoretical models of Lorentz violation in the current
universe (and experimental constraints thereon) have been extensively
studied in recent years
\cite{{Zee:1981sy},{Carroll:1989vb},{Coleman:1997xq},{Colladay:1998fq},{Soda:2006wr}}.
Here we are specifically concerned with the possibility that
rotational invariance may have been broken during inflation by an
effect that has subsequently disappeared, and study the effects of
such breaking on CMB anisotropies.  It is possible that such an effect
has already been detected, in the form of the ``Axis of Evil," an
apparent alignment of the CMB multipoles on very large scales
\cite{{deOliveira-Costa:2003pu},{Eriksen:2003db},{Copi:2003kt},{Schwarz:2004gk},{Hansen:2004vq},{Prunet:2004zy},{Eriksen:2004iu},{Land:2005ad},{Jaffe:2005pw},{Copi:2005ff},{Land:2005cg},{Bernui:2006ft},{Wx},{Abramo:2006gw},{Magueijo:2006we},{Hajian:2006ud},{Park:2006dv},{Huterer:2006nb},{Vx},{Land:2006bn},{Eriksen:2007pc}}.
Although its statistical significance is hard to quantify, a variety
of models have been put forward to explain this phenomenon
\cite{{Berera:2003tf},{Donoghue:2004gu},{Gordon:2005ai},{Armendariz-Picon:2005jh},{Buniy:2005qm},{Cresswell:2005sh},{Alexander:2006mt},{Battye:2006mb},{Campanelli:2006vb},{Gumrukcuoglu:2006xj}}.
Our aim is not to construct a model contrived to explain the currently
observed large-scale anomalies, but rather to make robust
predictions for the observable consequences of a preferred direction
during inflation, allowing observations to put constraints on its
magnitude.

The power spectrum $P(k)$ for the primordial density perturbations $ \delta({\bf k})$ is defined by
\begin{equation}
\label{powera}
\langle \delta({\bf k})  \delta^*({\bf q})\rangle= P(k)\delta^3({\bf k}-{\bf q}).
\end{equation}
The Dirac delta function in Eq.~(\ref{powera}) implies that modes with different wavenumbers are uncoupled, and is a consequence of translational invariance during the inflationary era. On the other hand, the fact that the power spectrum $P(k)$ only depends on the magnitude of the vector ${\bf k}$ is a consequence of rotational invariance. Suppose that during the inflationary era rotational invariance is broken by the presence of a small vector that points in the direction of a unit vector ${\bf n}$. Assuming a parity ${\bf k} \rightarrow -{\bf k}$ symmetry, the leading effect of the violation of rotational invariance changes the most general form of the power spectrum from $P(k)$ to $P'({\bf k})$, where
\begin{equation}
\label{power}
P'({\bf k})=P(k)\left(1+g(k)(\hat{\bf k}\cdot {\bf n})^2   \right).
\end{equation}
Here $\hat {\bf k}$ is the unit vector along the direction of $\bf k$ and we are neglecting higher powers of ${\hat {\bf k}}\cdot {\bf n}$ since they will be suppressed by more powers of the magnitude of the small vector that breaks rotational invariance.  (Effects of a \emph{timelike} vector on inflationary perturbations have also been studied \cite{Lim:2004js}.)

Towards the end of the inflationary era, the physical wavelengths that correspond to scales of astrophysical interest are large compared with the inverse Hubble constant during inflation or any of the dimensionful particle-physics quantities that might be  relevant during inflation. The same naturalness arguments that lead to the scale-invariant Harrison-Zeldovich spectrum (i.e., primordial $P(k) \propto 1/k^3$) imply that $g(k)$ in Eq.~(\ref{power}) should be independent of $k$. Assuming that $g(k)$ is a $k$-independent constant $g_*$ over the scales of astrophysical interest, we arrive at
\begin{equation}
\label{power1}
P'({\bf k})=P(k)\left(1+g_*({\hat {\bf k}}\cdot {\bf n})^2  \right) .
\end{equation}
This is the form of the primordial power spectrum that takes into account the leading effects of the violation of rotational invariance by a small vector in the inflationary era that points in the direction $\bf n$. In the next section we discuss the implications of the power spectrum in Eqs.~(\ref{power}) and~(\ref{power1}) for the anisotropy of the microwave background radiation. The breaking of rotational invariance gives rise to correlations between multipole moments that would normally vanish and also alters the predictions for the usual multipole moment correlations. In section three we discuss a simple model that realizes the form of the primordial power spectrum in Eq.~(\ref{power1}). Concluding remarks are given in section four.

\section{Microwave Background}

We are interested in a quantitative understanding of how the substitution, $P(k) \rightarrow P'({\bf k})$, changes the prediction for the microwave background anisotropy $\Delta T/T$.  The multipole moments are defined by
\begin{equation}
\label{def1}
a_{l m}=\int {\rm d} \Omega_{\bf e} (Y_l^m({\bf e}))^*{\Delta T \over T}({\bf e}).
\end{equation}
The anisotropy of the microwave background temperature $T$ along the direction of the unit vector $\bf e$ is related to the primordial fluctuations by
\begin{equation}
\label{def2}
{\Delta T \over T}({\bf e})=\int{\rm d}{\bf k} \sum_{l} \left({2l+1
\over 4 \pi}\right) (-i)^l P_l({\hat {\bf k}}\cdot {\bf e})\delta ({\bf k})\Theta_l(k),
\end{equation}
where $P_l$ is the Legendre polynomial of order $l$ and $\Theta_l(k)$ is a function of the magnitude of the wave-vector ${\bf k}$ that includes, for example, the effects of the transfer function. It can only depend on the magnitude of the wave-vector since the dynamics after the inflationary era is assumed to be rotationally invariant. 

We would like to compute the expectation values $\langle a_{lm}a_{l'm'}^* \rangle$ to first order in the small quantity $g(k)$ that characterizes the primordial violation of rotational invariance. We write
\begin{equation}
\langle a_{lm}a_{l'm'}^* \rangle=\langle a_{lm} a_{l'm'}^* \rangle_0+ \Delta(lm;l'm') ,
\end{equation}
where the subscript $0$ denotes the usual rotationally invariant piece,
\begin{equation}
\langle a_{lm} a_{l'm'}^* \rangle_0=\delta_{l l'}\delta_{m m'}\int_0^{\infty} {\rm d}k k^2 P(k)\Theta_l(k)^2.
\end{equation}
It is useful to introduce the ``spherical" components of the unit vector ${\bf n}$ that defines the preferred direction for rotational non-invariance,
\begin{equation}
n_+=-\left( {n_x-in_y \over \sqrt{2}} \right),~n_-=\left( {n_x+in_y
\over \sqrt{2}} \right),~n_0=n_z. 
\end{equation} 
In terms of these components the unit norm condition becomes $n_0^2-2n_+n_-=1$. Note that we do not assume that the preferred direction ${\bf n}$ coincides with the ${\hat{\bf z}}$ axis of the coordinate system used to parameterize the microwave sky (i.e., that $n_+=n_-=0$).  Expressions analogous to ours have been derived by G\"umr\"uk\c{c}\"uo\u{g}lu et al. \cite{Gumrukcuoglu:2006xj} under the assumption that these two directions are coincident; see also \cite{Hajian:2006ud}.

Using the identity
\begin{equation}
P_l({\hat {\bf k}}\cdot {\bf e})={4 \pi \over 2l+1}\sum_{m=-l}^{l} Y_l^m({\bf e}) (Y_l^m({\hat {\bf k}}))^*,
\end{equation}
it is straightforward to express the sought-after perturbation as
\begin{equation}
\label{Delta}
\Delta(lm;l'm')=(-i)^{l-l'} \xi_{lm;l'm'}\int_0^{\infty} {\rm d}k k^2 P(k)g(k)\Theta_l(k)\Theta_{l'}(k),
\end{equation}
where
\begin{eqnarray}
\xi_{lm;l'm'}&=&{4 \pi \over 3}\int {\rm d}\Omega_{\bf k}(Y_l^m(\hat{\bf k}))^* Y_{l'}^{m'}(\hat{\bf k})  \\
&&\times\left(n_+Y_1^1(\hat{\bf k})+n_-Y_1^{-1}(\hat{\bf k})+n_0Y_1^0(\hat{\bf k}) \right)^2 \nonumber 
\end{eqnarray}
(we use the Condon-Shortley phase convention for the spherical harmonics; see \cite{Arfken}).

The integral in (\ref{Delta}) encodes information about the power spectrum and the transfer function, as well as the scale-dependence of the preferred-direction effect, while the constants $\xi_{lm;l'm'}$ are purely geometric.  The integration over solid angles is straightforward to perform. It is convenient to decompose the $\xi_{lm;l'm'}$ into coefficients of the quadratic quantities $n_in_j$, via
\begin{eqnarray}
\label{xiexpansion}
\xi_{lm;l'm'}&=&n_+^2\xi_{lm;l'm'}^{++}+n_-^2\xi_{lm;l'm'}^{--}+2n_+n_-\xi_{lm;l'm'}^{+-} \\
&&+2n_+n_0\xi_{lm;l'm'}^{+0}+2n_-n_0\xi_{lm;l'm'}^{-0}+n_0^2\xi_{lm;l'm'}^{00}.\nonumber
\end{eqnarray}
These coefficients are then given by the following expressions:
\begin{widetext}
\begin{eqnarray}
\xi_{lm;l'm'}^{--}&=&-\delta_{m',m+2} \left[\delta_{l',l}{\sqrt{
(l^2-(m+1)^2)(l+m+2)(l-m)}\over (2l+3)(2l-1)}-{1 \over
2}\delta_{l',l+2}{\sqrt{ {(l+m+1)(l+m+2)(l+m+3)(l+m+4)\over
(2l+1)(2l+3)^2(2l+5)}}}\right. \nonumber \\ 
&& \left. -{1 \over 2}\delta_{l',l-2}{\sqrt{ {(l-m)(l-m-1)(l-m-2)(l-m-3)\over (2l+1)(2l-1)^2(2l-3)}}}\right] ,\nonumber \\
\xi_{lm;l'm'}^{++}&=&\xi_{l'm';lm}^{--},\nonumber \\
\xi_{lm;l'm'}^{+-}&=&{1 \over 2}\delta_{m',m}\left[-2 \, \delta_{l',l}
\frac{(-1+l+l^2+m^2)}{(2l-1)(2l+3)} \right. +
 \delta_{l',l+2}\sqrt{\frac{((l+1)^2-m^2)((l+2)^2-m^2)}{(2l+1)(2l+3)^2(2l+5)}}
\nonumber \\
&& +\left.
\delta_{l',l-2}\sqrt{\frac{(l^2-m^2)((l-1)^2-m^2)}{(2l-3)(2l-1)^2(2l+1)}}\right],
\nonumber \\
\xi_{lm;l'm'}^{-0}&=&-\frac{1}{\sqrt{2}}\delta_{m',m+1}\left[
\delta_{l',l} \frac{(2m+1)\sqrt{(l+m+1)(l-m)}}{(2l-1)(2l+3)} + 
\delta_{l',l+2}\sqrt{\frac{((l+1)^2-m^2)(l+m+2)(l+m+3)}{(2l+1)(2l+3)^2(2l+5)}}\right.
\nonumber \\
&&- \left. 
\delta_{l',l-2}\sqrt{\frac{(l^2-m^2)(l-m-1)(l-m-2)}{(2l-3)(2l-1)^2(2l+1)}}
\right], \nonumber\\
\xi_{lm;l'm'}^{+0}&=&-\xi_{l'm';lm}^{-0},\nonumber \\
\xi_{lm;l'm'}^{00}&=&\delta_{m,m'} \left[ \delta_{l,l'} \frac{(2l^2+2l-2m^2-1)}{(2l-1)(2l+3)} +\delta_{l',l+2}
\sqrt{\frac{((l+1)^2-m^2)((l+2)^2-m^2)}{(2l+1)(2l+3)^2(2l+5)}}
\right. \nonumber \\ 
&&\left. +\delta_{l',l-2} \sqrt{\frac{(l^2-m^2)((l-1)^2-m^2)}{(2l-3)(2l-1)^2(2l+1))}} \right].
\label{thebigformula}
\end{eqnarray}
\end{widetext}

The formulas (\ref{xiexpansion},\ref{thebigformula}) are explicit expressions for the geometrical part of the perturbation (\ref{Delta}).  As we mentioned in the introduction, it is natural to imagine that the violation of rotational invariance is approximately scale invariant, which implies that it is a good approximation to set $g(k)=g_*$, a constant.  If we define polar coordinates
$\theta_*, \phi_*$ for the preferred direction, 
\begin{equation}
n_x={\rm
sin}\theta_*{\rm cos}\phi_* \, , \,
n_y={\rm sin}\theta_*{\rm sin}\phi_* \, , \,
n_z={\rm cos}\theta_* \, ,
\end{equation}
these expressions can be compared directly with observations to constrain the three parameters $(g_*, \theta_*, \phi_*)$.

When $g(k)=g_*$, a simplification occurs for $l=l'$ and $m=m'$, as the dependence on the power
spectrum for the terms that violate rotational invariance $\Delta(lm;lm)$ is the same
as the rotationally-invariant part $\langle a_{lm} a_{lm}^* \rangle_0$.
We can then find a simple expression for their ratio,
\begin{eqnarray}
\label{wow}
{\Delta(lm;lm) \over \langle a_{lm} a_{lm}^* \rangle_0}
&=&{g_* \over 2}\Bigg[ {\rm sin}^2\theta_*+ \\
&&  (3{\rm cos}^2\theta_*-1)\left({2l^2+2l-2m^2-1 \over (2l-1)(2l+3)}\right)\Bigg].\nonumber 
\end{eqnarray}
For large multipoles, $l \gg 1$, and for the magnitude of $m$ of the
order of $l$, this expression simplifies to
\begin{equation}
\label{wow1}
{\Delta(lm;lm) \over \langle a_{lm}a_{lm}^* \rangle_0}={g_* \over 4}\Bigg[ 1+{\rm cos}^2\theta_*-  (3{\rm cos}^2\theta_*-1){m^2 \over l^2}\Bigg].
\end{equation}

\section{Inflation Model with a Preferred Direction}

It is interesting to see how the rotationally non-invariant power spectrum in Eq.~(\ref{power1}) can arise in an explicit model of anisotropic inflation.  We will assume that, during most of the inflationary era, rotational invariance is broken by a spacelike four-vector $u^\mu$ with invariant length
\begin{equation}
\label{fix}
g_{\mu \nu}u^{\mu}u^{\nu}= m^2.
\end{equation}
We will consider the effect of the energy-momentum tensor associated with this vector on the expansion of the universe during inflation, ignoring direct couplings of $u^\mu$ to other fields.  Gravitational effects of dynamical Lorentz-violating vector fields have been considered previously in the literature \cite{{Kostelecky:1989jw},{Jacobson:2000xp},{Jacobson:2004ts},{Carroll:2004ai},{Eling:2003rd},{Kostelecky:2003fs}}.

We assume that the four-vector $u^\mu$ is non-zero only during the
time interval $0<t<t_*$, where $t_*$ is the end of inflation, so that
the dynamics is rotationally invariant during reheating and
thereafter. During the time interval $0<t<t_*$, the dynamics of
interest is governed by the action 
\begin{equation}
\label{dyn}
S=\int {\rm d}^4x {\sqrt -g}\left({1 \over 16 \pi G} R-\rho_{\Lambda} +{\cal L}_u +{\cal L}_{\chi}\right),
\end{equation}
where
\begin{equation}
{\cal L}_{\chi} =-{1 \over 2}g^{\mu \nu}\partial_{\mu}\chi \partial_{\nu}\chi
\end{equation}
and
\begin{eqnarray}
\label{ulag}
  {\cal{L}}_u &=& -\beta_1 \nabla^\mu u^\sigma \nabla_\mu u_\sigma - \beta_2
  (\nabla_\mu u^\mu)^2  \\ \nonumber
&& - \beta_3 \nabla^\mu u^\sigma \nabla_\sigma
u_\mu + \lambda(u^{\mu} u_{\mu} - m^2)\ .
\end{eqnarray}

Here $\lambda$ is a Lagrange multiplier that enforces the constraint
(\ref{fix}). Quantum fluctuations in the massless scalar field $\chi$
are assumed to dominate the density perturbations via the DGZK
mechanism~\cite{DGZK}. 
In that case we need simply calculate the fluctuations in $\chi$,
without worrying about the behavior of the inflationary potential.  

We approximate the inflaton energy density as a constant, modeling the effects of the inflaton field by a vacuum energy $\rho_{\Lambda}$ in Eq.~(\ref{dyn}). The inflationary spacetime is taken of the form
\begin{equation}
\label{coord}
{\rm d}s^2=-{\rm d}t^2+a(t)^2{\rm d}{\bf x}_{\perp}^2+b(t)^2{\rm d}z^2
\end{equation}
since we have chosen the four-vector to be aligned along the $z$-axis
direction,
\begin{equation}
u^0=0~~,u^x=0~~,u^y=0~~,u^z={m \over b(t)}.
\label{uvector}
\end{equation}
The energy-momentum tensor for $u^\mu$ derived from (\ref{ulag}) is \cite{Carroll:2004ai}
\begin{eqnarray}
\label{tmunu}
T_{\mu\nu}^{(u)}&=&2\beta_1(\nabla_{\mu} u^{\rho}\nabla_{\nu} u_{\rho}-\nabla^{\rho} u_{\mu} \nabla_{\rho} u_{\nu})  \nonumber \\
&&-2[\nabla_{\rho}(u_{(\mu}J^{\rho}{}_{\nu)})+\nabla_{\rho}(u^{\rho} J_{(\mu\nu)})-\nabla_{\rho}(u_{(\mu}J_{\nu)}{}^{\rho})] \nonumber \\
&&+2m^{-2}u_{\sigma}\nabla_{\rho} J^{\rho\sigma}u_{\mu} u_{\nu}+g_{\mu\nu}{\cal{L}}_u ,
\end{eqnarray}
where $J^{\mu}{}_{\sigma}$ is the current tensor,
\begin{equation}
J^{\mu}{}_{\sigma}= -\beta_1 \nabla^{\mu} u_{\sigma}-\beta_2
\, \delta^{\mu}_{\sigma} \, \nabla_{\rho} u^{\rho} -\beta_3 \nabla_{\sigma} u^{\mu}.
\nonumber
\end{equation}

Given (\ref{uvector}) and (\ref{tmunu}), the nonvanishing components
of the stress tensor are
\begin{eqnarray}
T_{00}^{(u)} &=&  \beta_1 m^2\left({\dot{b}\over b}\right)^2
\nonumber\\
T_{xx}^{(u)}&=&T_{yy} = \beta_1 m^2 a^2 \left({\dot{b}\over b}\right)^2
\nonumber\\
T_{zz}^{(u)} &=&  \beta_1 m^2 \left(  \dot{b}^2 -2 \ddot{b} b -4 {\dot{a}\dot{b} b \over a} \right).
\end{eqnarray}
Note that the components of the energy momentum tensor in our chosen
background are independent of $\beta_2$ and $\beta_3$. 

Solving Einstein's equation during the time interval $0<t<t_*$, with initial conditions $a(0)=1$ and $b(0)=1$, gives
\begin{equation}
a(t)=e^{H_a t},~~~b(t)=e^{H_b t},
\end{equation}
where
\begin{eqnarray}
&&H_a= {{\dot a}\over a} = H_b(1+16 \pi G  \beta_1 m^2),\nonumber  \\
&& H_b= {{\dot b}\over b} 
={{\sqrt{ 8 \pi G \rho_{\Lambda} \over (1+8 \pi G  \beta_1 m^2)(3+32
\pi G  \beta_1 m^2)}}} \,.
\end{eqnarray}

According to the cosmic no-hair theorem, initially expanding
homogeneous cosmological models in the presence of a positive
cosmological constant will rapidly approach a de~Sitter solution, if
the other matter fields obey the dominant and strong energy conditions
\cite{wald}.  Our specific model violates these conditions.
Nevertheless, for $\beta_3 = - \beta_1$ and $\beta_2=0$ the kinetic term
for fluctuations about our background has the form of a field strength
tensor squared and so is ghost free. We therefore expect the
configuration to be stable with respect to small fluctuations.

It will turn out to be convenient to refer to a fictitious isotropic metric,
\begin{equation}
{\rm d}{\bar s}^2=-{\rm d}t^2+\bar{a}(t)^2[{\rm d}x^2 + {\rm d}y^2 + {\rm d}z^2],
\end{equation}
in which the scale factor expands exponentially
\begin{equation}
{\bar a}(t)=e^{\bar{H}t}
\end{equation}
with an ``average'' Hubble parameter,
\begin{equation}
\bar{H}={1\over 3}(2H_a+H_b). 
\end{equation}
Deviations from isotropy can be parameterized by
\begin{equation}
  \epsilon_H = \frac{2}{3}\left(\frac{H_b - H_a}{\bar H}\right) ,
\end{equation}
where the $2/3$ will become useful later.
We work in the limit $N_*|\epsilon_H| <<1$, where $N_*={\bar H} t_*$ is the number of $e$-foldings during the time when the four-vector $u^\mu$ is non-zero.  This assures that the violation of rotational invariance due to the anisotropic expansion is always a small perturbation.

We need to compute the correlation function $\langle \chi({\bf x},t)
\chi({\bf y},t) \rangle$. Treating $\epsilon_H$ as a small
perturbation, we find that to first order in this quantity we obtain
(see for example ref. \cite{Weinberg:2005vy})
\begin{eqnarray}
\langle \chi({\bf x},t) \chi({\bf y},t) \rangle &\simeq&
\langle \chi_I({\bf x},t)\chi_I({\bf y},t) \rangle\\
&& +i\int_0^t{\rm d}t'\langle [H_I(t'),\chi_I({\bf x},t) \chi_I({\bf y},t) ]\rangle.  \nonumber 
\label{2pcf}
\end{eqnarray}
Here the interaction-picture Hamiltonian $H_I(t)$ is given by
\begin{eqnarray}
H_I(t)&=&\int {\rm d}^3x{1 \over 2} \left[ (b(t)-{\bar a}(t)) \left({\rm d} \chi_I \over {d{ \bf x}_{\perp}}\right)^2 \right.\nonumber \\
&&+\left.\left({a(t)^2 \over b(t)}-{\bar a}(t) \right)\left({\rm d} \chi_I \over {dx^3}\right)^2 \right].
\end{eqnarray}
The interaction-picture (i.e. free) field obeys the rotationally-invariant equation of motion,
\begin{equation}
\label{evolution}
{{\rm d}^2 \chi_I \over {\rm d} t^2}+3 {\bar H} {{\rm d} \chi_I \over {\rm d} t}-{1 \over {\bar a}(t)^2}{{\rm d}^2\chi_I \over {\rm d}{\bf x}^2}=0.
\end{equation}

We can write the two-point correlation function (\ref{2pcf}) in terms of Fourier transforms as
\begin{eqnarray}
\langle \chi({\bf x},t) \chi({\bf y},t) \rangle &=&\int {{\rm d}^3k \over (2\pi)^3} e^{-i{\bf k}\cdot ({\bf x}-{\bf y})}\left[P(k) \right. \nonumber \\
&&+ \left.(\hat{\bf k}\cdot {\bf n})^2\Delta P(k)\right].
\end{eqnarray}
Converting to the conformal time of the isotropic metric,
\begin{equation}
\tau=-{ 1\over {\bar H}}e^{-{\bar H} t},
\end{equation}
and expanding in $\epsilon_H$, we find that $P(k) \simeq |\chi_k^{(0)}(\tau)|^2$, and
\begin{eqnarray}
&&\Delta P(k) \simeq 3ik^2\epsilon_H\int_{-1/{\bar H}}^{\tau} {\rm d}\tau'
\left(- {1 \over {\bar H} \tau'}\right)^2 \\
&&\times{\rm log}(-{\bar H} \tau')\left[(\chi_k^{(0)}(\tau')\chi_k^{(0)}(\tau)^*)^2-(\chi_k^{(0)}(\tau')^*\chi_k^{(0)}(\tau))^2\right] \nonumber,
\end{eqnarray}
where 
\begin{equation}
\chi_k^{(0)}(\tau)={{\bar H} \over \sqrt{2k}}e^{-ik\tau}\left[ \tau-{i \over k} \right].
\end{equation}
We assume that the modes $k$ of astrophysical interest have
wavelenghts much smaller than the Hubble radius at the beginning of
inflation, which in our normalization implies $k>>{\bar H}$.  They
cross the horizon around sixty $e$-foldings before
the end of inflation  (which we take to occur at about $t_*$). Taking
$|k \tau| <<1$, we find that 
\begin{equation}
\label{deltaP}
\Delta P(k)\simeq \frac{9}{4}\epsilon_H{{\bar H}^2 \over  k^3}{\rm log}(k/{\bar H}),
\end{equation}
where we have neglected contributions not enhanced by the large logarithm. 

There is another way to derive Eq.~({\ref{deltaP}). For modes with wavenumbers  along the $\hat {\bf z}$ direction or perpendicular to this direction, the Fourier transform of the two point function $\langle \chi({\bf x},t) \chi({\bf y},t) \rangle$ can be found exactly without resorting to perturbation theory. For example, modes $\chi_k$ with ${\bf k}=k \hat{\bf z}$ (wavevectors parallel to the preferred direction) obey the differential equation
\begin{equation}
\label{evolution1}
{{\rm d}^2 \chi_k \over {\rm d} t^2}+3 {\bar H} {{\rm d} \chi_k \over {\rm d} t}+{k^2 \over b(t)^2}\chi_k=0.
\end{equation}
The canonical commutation relations imply that $\chi_k$ satisfies the normalization condition,
\begin{equation}
\left({{\rm d}\chi_k(\tau) \over {\rm d}\tau}\right)\chi_k(\tau)^*-\left({{\rm d}\chi_k(\tau)^*\over {\rm d}\tau}\right)\chi_k(\tau)=-i({\bar H} \tau)^2.
\end{equation} 
We find that the properly normalized solution to Eq.~(\ref{evolution1}) is
\begin{equation}
\chi_k(\tau)={{\bar H} \sqrt{\pi} \tau^{3/2}\over 2 \sqrt{1+\epsilon_H}}H_{\nu}^{(2)}
\left({(k/{\bar H})^{-\epsilon_H} (k \tau)^{1+\epsilon_H} \over 1+\epsilon_H }\right),
\end{equation}
where $H_\nu^{(2)}$ is a Hankel function, and
\begin{equation}
\nu={3 \over 2 + 2\epsilon_H}.
\end{equation}
The contribution to the Fourier transform of the two point $\chi$ correlation for a mode along the $\hat{\bf z}$ direction is $|\chi_k(\tau)|^2$. For small $\epsilon_H$ and $|k \tau|$ and large $k/{\bar H}$, this becomes
\begin{equation}
|\chi_k(\tau)|^2 \simeq {{\bar H}^2 \over 2 k^3}\left(1 +3\epsilon_H {\rm log}(k/{\bar H})\right).
\end{equation}
Here we have neglected terms linear in $\epsilon_H$ that are not enhanced by the large logarithm. Combining this result with a similar analysis for modes perpendicular to the $\hat{\bf z}$ direction reproduces the result in Eq.~(\ref{deltaP}). 

Finally we note that the density perturbation power spectrum is defined by a Fourier transform with respect to coordinates where physical laws have manifest rotational invariance. However at time $t=t_*$ the coordinates in Eq.~(\ref{coord}) do not exhibit manifest rotational invariance, due to the difference between $a(t_*)$ and $b(t_*)$. 
Rescaling coordinates, $z \rightarrow z({\bar a}(t_*)/b(t_*))$ and
${\bf x}_{\perp} \rightarrow {\bf x}_{\perp}({\bar a}(t_*)/a(t_*))$, we find that the function $g(k)$  characterizing the rotationally non-invariant part of the power spectrum for the primordial density perturbations is
\begin{eqnarray}
\label{g}
g(k)&=&\frac{9}{2} \epsilon_H ({\rm log}(k/{\bar H})-N_*) \nonumber \\
&&=\frac{9}{2} \epsilon_H {\rm log}(q(t_*)/{\bar H}),
\end{eqnarray}
where the term proportional to $N_*$ comes from the rescaling of
coordinates and $q(t_*)=k/{\bar a}(t_*)$ is the physical wavelength of
the mode of interest at the end of inflation. 

The logarithm in (\ref{g}) is actually nearly constant over values of
$q(t_*)$ of astrophysical interest.  The range of $q(t_*)$ probed by
CMB measurements is about a factor of $10^3$, so $\log(q(t_*)/{\bar H})$ changes by roughly
$7$.  But the modes of cosmological interest cross the deSitter horizon around
60 $e$-foldings before the end of inflation. So $|\log(q(t_*)/{\bar
H})|$ is approximately 60. Hence, in
this model $g(k)$ varies by about $10\%$ over the range of modes of
cosmological interest and our general expectation that setting $g(k)=g_{*}$
is a reasonable approximation has been confirmed. 

For simplicity in this analysis we neglected terms that directly couple $u^\mu$ to $\chi$. For example we could have added the term  $u^{\mu}u^{\nu}\partial_{\mu}\chi \partial_{\nu} \chi/M^2$ to the Lagrange density. It is easy to see that this gives an additional scale invariant contribution, $3 m^2 /M^2$, to $g(k)$.

\section{Concluding Remarks}

We have investigated the possibility that rotational invariance may have been explicitly broken during inflation by an effect that has disappeared in the later universe.  The observed CMB temperature anisotropies provide a direct window onto the physics of the inflationary era, and therefore offer a unique opportunity for constraining (and discovering) new phenomena at high scales.  Our aim has been to investigate the generic predictions we expect from the presence of a preferred direction during inflation.

If rotational invariance is violated during inflation, it is natural for the effects of such a violation to show up in a scale-invariant way, just as the amplitude of the perturbations themselves are approximately scale-invariant.  Under that assumption, we derive a powerful set of predictions for the expectation values $\langle a_{lm} a_{l'm'}^*\rangle$ that depend on only three parameters:  a single amplitude $g_*$, and a direction on the sky defined by a unit vector ${\bf n}$.  Investigation of a simple model confirms the approximate scale-independence of this effect.  The resulting expressions (\ref{Delta},\ref{xiexpansion},\ref{thebigformula}) can be directly compared with observations to probe the existence of small Lorentz-violating effects in the very early universe.

\section*{Acknowledgments}

We are thankful for interesting discussions with Jonathan R. Pritchard at the beginning of this work, Suz Tolwinski before it even began, and Chris Gordon and Dragan Huterer near its completion.  This research was supported in part by the U.S. Department of Energy and by the David and Lucile Packard Foundation.

\end{document}